\documentclass[preprint,titan,amsmath,amssymb,floatfix,showpacs]{revtex4}
\usepackage{graphicx}% Include figure files
\usepackage{bm}% bold mat
\usepackage{amssymb}
\usepackage{color}
\usepackage{epsfig}
\usepackage{subfigure}

\newcommand{\be}{\begin{equation}}
\newcommand{\ee}{\end{equation}}

\begin{document}
%
%\keywords{\emph{1/f }noise, electron glasses, slow relaxations,
%aging.}
%\subjclass[pacs]{71.23.Cq, 05.40.-a, 73.50.-h} % up to three, separated by commas
%
%% 05.40.-a Fluctuation phenomena, random processes, noise, and Brownian motion
%% 72.80.Ng Disorder solids
%% 72.80.Sk insulators
%% 71.23.Cq Amorphous semiconductors, metallic glasses, glasses
%% 72.20.Ee Mobility edges; hopping transport
%% 73.50.-h Electronic transport phenomena in thin films
%
%
%\makeatletter
%%\AddToShipoutPicture{%
%%            \setlength{\@tempdimb}{.5\paperwidth}%
%%            \setlength{\@tempdimc}{.5\paperheight}%
%%            \setlength{\unitlength}{1pt}%
%%            \put(\strip@pt\@tempdimb,\strip@pt\@tempdimc){%
%%        \makebox(0,0){\rotatebox{45}{\textcolor[gray]{0.75}%
%%        {\fontsize{6cm}{6cm}\selectfont{DRAFT}}}}%
%%            }%
%%} \makeatother

\title{\emph{1/f }noise and slow relaxations in glasses}

\author{Ariel Amir, Yuval Oreg, Yoseph Imry}

\affiliation { Department of Condensed Matter Physics, Weizmann
Institute of Science, Rehovot, 76100, Israel\\}
%
%\author[Ariel Amir]{Ariel Amir \footnote{Corresponding
%     author \quad E-mail: {\sf ariel.amir@weizmann.ac.il}}}
%    % Phone: +972\8\934\2088,
%    % Fax: +972\8\934\4477}
%\address[]{Department of Condensed Matter Physics, Weizmann
%Institute of Science, Rehovot, 76100, Israel} %%
%\author[Yuval Oreg]{Yuval Oreg}
%
%\author[Yoseph Imry]{Yoseph Imry}

\begin{abstract}
Recently we have shown that slow relaxations in the electron glass
system can be understood in terms of the spectrum of a matrix
describing the relaxation of the system close to a metastable state.
The model focused on the electron glass system, but its generality
was demonstrated on various other examples. Here, we study the noise
spectrum in the same framework. We obtain a remarkable relation
between the spectrum of relaxation rates $\lambda$ described by the
distribution function $P(\lambda) \sim 1/\lambda$ and the $1/f$
noise in the fluctuating occupancies of the localized electronic
sites. This noise can be observed using local capacitance
measurements. We confirm our analytic results using numerics, and
also show how the Onsager symmetry is fulfilled in the system.
\end{abstract}

\maketitle

\section{Introduction}
\label{sect1}

\emph{1/f} noise is ubiquitous and has been observed in a variety of
physical systems, such as metals, semiconductors and superconductors
\cite{dutta, Weissman}, as well as biological and economic systems
\cite{heartbeat, economic}. Systems displaying slow relaxations and
aging are also abundant, and a broad range of materials, such as
spin glasses \cite{spinglass}, structural glasses
\cite{structural_ludwig}, electron glasses \cite{zvi} and biological
systems \cite{thompson} have been shown to exhibit these properties.

There has been much interest in studying \emph{1/f} noise in
electron glasses, both experimentally and theoretically
\cite{{voss}, {shklovskii}, {kogan1},
{kozub},{kogan2},{shklovskii2},{shklovskii3},{shtengel},{galperin},massey, mccammon, kar, zvi_exp1}. Yu
\cite{yu2} has motivated the study of noise in electron glasses,
particularly the study of the 'second noise' (describing the
fluctuations in the noise spectrum), as a tool to distinguish
between different glassy models.

In a recent work \cite{amir_aging}, we have proposed a model showing
that the slow relaxations observed in the electrons glasses can be
understood in terms of an underlying $1/\lambda$ distribution of
relaxation rates, which arise naturally as the eigenvalues of a
certain class of random matrices. It is the purpose of this work to
show that the same model predicts a \emph{1/f} noise spectrum for
the site occupancies, arising directly from the $1/\lambda$
distribution. This establishes a remarkable relation between
\emph{1/f} noise and the universal slow relaxations. We study in
detail the electron glass model, but the established connection
between the two should be much more general than this specific
system of interest, since  the same equations can be used to
describe a broad range of systems.

The structure of the manuscript is as follows: we first define the
model, and discuss briefly the local mean-field approximation used.
We then review the previously derived results for the spectrum of
the relaxation rates, and set up the equations describing the
near-equilibrium fluctuations. We show that the Onsager symmetry is
obeyed in the system, resulting in the specific form of the
non-hermitian relaxation matrix, and show that using \emph{Onsager's
regression hypothesis} we can relate the noise spectrum with the
$1/\lambda$ relaxation spectrum. Finally, we show numerical support
for the calculation, and propose an experimental test for it.

\section{Definition of the model and the local mean-field approximation}
\label{sect2}

The model typically used to study electron glasses addresses a set
of states, which are assumed to be Anderson localized due to the
strong disorder present. While many studies have been performed,
usually for numerical convenience, on a lattice, we choose to work
with localized states positioned randomly in space, which we find
more realistic. The number of electrons $M$ is smaller than the
number of sites $N$. The states' on-site energies contain disorder,
of typical magnitude $W$. Due to the localization, the Coulomb
interactions between the electrons are not screened. They can not be
neglected in the analysis and are not neglected within the model.
The electrons are coupled to a phonon bath, assumed at thermal
equilibrium. Details of the model are found in
[\onlinecite{amir_glass}], where we use a local mean-field
approximation \cite{coulomb_gap_mean_field} to study this model . We
should emphasize that within this approximation the on-site and
spatial disorder are not averaged over. We find that the
approximation captures the well-known results for the Coulomb gap
\cite{efros2, efros} in the density-of-states (DOS), a soft gap
emerging near the Fermi-energy as a result of the Coulomb
interactions. It was also used to describe the slow relaxations
experimentally observed in an aging experiment \cite{amir_aging}. In
a later study we have shown that this model also describes the
transition in the hopping conductance, from Mott's variable range
hopping at higher temperatures to Efros-Shklovskii variable range
hopping below the crossover \cite{amir_VRH}.

An important point is that within the local mean-field
approximation, the occupation numbers are continuous variables, and
their dynamics is described by a coupled set of $N$ non-linear
differential equations. Metastable states are configurations of
occupation numbers for which the dynamics vanishes. In
[\onlinecite{amir_glass}] we show that one can understand the
dynamics near one of the many metastable states by linearizing the
equations of motion close to it. This led us to an equation for the
vector of deviations $\vec{\delta n}$ from the equilibrium
configuration in which the i'th site has the occupation $n^0_i$:

\be \frac{d \vec{\delta n}}{dt}= A \cdot \vec{\delta n}, \label
{dynamics} \ee

where the explicit form of the non-diagonal elements of $A$ is:

\be { A_{ij}= \gamma_{ij}\frac{1}{n_j^0 (1-n_j^0)} -\sum_{k \neq j,
i} \frac{e^2 \gamma_{ik}}{T}( \frac{1}{r_{ij}}- \frac{1}{r_{jk}})},
\label{realistic} \ee

with $\gamma_{ij}$ the equilibrium current from site $i$ to site $j$
(detailed balance is equivalent to $\gamma$ being a symmetric
matrix). From particle number conservation it follows that the sum
of each column of $A$ vanishes. Therefore the diagonal element of
the matrix is minus the sum of the rest of the elements in the
column, i.e., $A_{ii}=-\sum_{j \neq i} A_{ij}$.

Notice that exactly at the metastable state there would be
absolutely no dynamics (or fluctuations) within this equation, even
though it describes a system at \emph{finite} temperature. This is
clearly not physical, and in subsection \ref{langevin} we add a
necessary Langevin noise term, and determine its magnitude from
thermodynamic considerations.

\section{Known results for the spectrum}
\label{sect3}

It can be shown that the eigenvalues of the matrix $A$ are real and
negative. For a proof that the eigenvalues are real see subsection
\ref{formofspectrum}, and for the negativity property see Ref. [17].
In [\onlinecite{amir_glass}] we have given numerical evidence that
the model described in section \ref{sect2} yields a spectrum which
is approximately:
\begin{equation}
 P(\lambda) \approx \frac{C}{\lambda} \;\;\;,\;\;\;
C=\frac{1}{\log\frac{\lambda_{\rm max}}{\lambda_{\rm min}}}
\label{eq:Plambda},
\end{equation}
where $\lambda_{\rm max}$ and $\lambda_{\rm min}$ are the upper and
lower cutoffs of the distribution.
 In a later work \cite{amir_RMT} we discuss a toy-model for this problem: we define a
class of random matrices whose sum of columns vanish, and where the
$i$,$j$'th element decays exponentially with the distance of points
$i$ and $j$, distributed randomly. Notice that these are exactly the
properties of the matrix $A$ described in section \ref{sect2}. The
merit of this simplification of the problem is that we manage to
find an analytical formula for the averaged spectrum of the
matrices, and verify the numerical result: up to logarithmic
corrections in dimensions higher than one, the spectrum follows the
$1 / \lambda$ spectrum in the low density regime. This is the
essential result we would use in the following, and which would lead
to the \emph{1/f }noise in the fluctuation of the occupation
numbers.

\section {Derivation of the noise spectrum}
\label{sect4}

 \emph{Onsager's regression hypothesis}, (see the Appendix for a derivation) states that the
\emph{equation of motion} of the correlation function $
\label{eq:defcor} \phi_{ij}(t)=\left\langle \delta n_i(t) \delta
n_j(t) \right\rangle$ is obtained by simply replacing $\delta
n_i(t)$ in the equation of motion (\ref{dynamics}) by the function
$\phi_{ij}(t)$: \be \label{eq:phi} \frac
{d\phi_{ij}(t)}{dt}=A_{ik}\phi_{kj}(t) .\ee

To find the correlation function one also needs the initial
conditions:

\be \phi_{ij}(0)=\left\langle \delta n_i(0) \delta n_j(0)
\right\rangle \equiv \beta^{-1}_{ij}. \ee

\subsection {Equilibrium correlations}
\label {Equilibrium correlations} By definition, the inverse of the
matrix $\beta$ describes the equilibrium correlations in the system.
We would now like to calculate $\beta$. Notice that generally
$\beta_{ij}=\beta_{ji}$ and that, for example, for a free Fermion
system we have $\beta^{-1}_{ij} = n_i(1-n_j) \delta_{ij}$.

In our case, we consider an \emph{interacting} Fermion system, and
it is not a-priori clear that we can use the above form. One might
argue that within the local mean-field approximation the problem is
essentially a non-interacting problem, with renormalized on-site
energies due to the Coulomb interactions. It is therefore reasonable
that the free Fermion result, not depending on the on-site energies,
would still be valid. However, we shall now show that there is an
additional term in $\beta$ arising from the interactions.  To do
this, we write the free energy\cite{ coulomb_gap_mean_field}:

\be F = \sum_i \epsilon_i \tilde{n}_i +\sum_{i \neq j} \frac{e^2
\tilde{n}_i \tilde{n}_j} {r_{ij}}+kT\sum_i(\frac{1}{2}+ \tilde{n}_i)
\log(\frac{1}{2}+ \tilde{n}_i)+(\frac{1}{2}-\tilde{n}_i) \log
(\frac{1}{2}-\tilde{n}_i),\ee  with $\tilde{n}_i \equiv n_i
-\frac{1}{2}$. The local mean-field equations can be obtained from
the minimalization condition $\frac{\partial
F}{\partial\tilde{n}_i}=0$. Notice that each site contains a
positive background charge of
$\frac{1}{2}$, to keep charge neutrality.%, so that the electronic
%charge per site is in fact $n_i=\frac{1}{2}+ \tilde{n}_i$.

Expanding $F$ near a metastable state (a local minima), we have:

\be F= F_0 + \frac{kT}{2}\sum_{i,j}\beta_{ij}\delta n_i \delta n_j
,\label {Fexpand}\ee

with:

\be \beta_{ij}= \delta_{ij}\frac{1}{n^0_i (1-n^0_i)}+ \frac{e^2}{kT
r_{ij}}.\label{beta} \ee

The correlation matrix is proportional to $\beta^{-1}$, since we
have a quadratic free energy:

\be \langle \delta n_i \delta n_j \rangle  =    \beta^{-1}_{ij}. \ee

The free fermion result is corrected for $i \neq j$ by the
interaction term. Notice that at low temperatures $n_i$ tend to 0 or
1 exponentially, and thus the matrix is nearly diagonal, i.e., the
interaction term is negligible.

The matrix $\beta$ has another physical meaning: if we introduce the
conjugate variables to $n_i$, the so-called thermodynamic forces
$\mu_i$, defined as $\frac{\partial F}{\partial n_i}$, we find that
\be \mu_i= \beta_{ik}n_k. \label{mu}\ee

\subsection {Onsager symmetry}
According to \emph{ Onsager's principle} (see Appendix) the kinetic
coefficients,
\begin{equation}
\label{eq:gamma} \gamma_{ij}=\sum_k A_{ik} \beta^{-1}_{kj}
\end{equation}
are symmetric:
\begin{equation}
 \gamma_{ij}=\gamma_{ji}.
\end{equation}

Indeed, Eqs. \ref{realistic} and \ref{beta} are related according
to:

\be A = \gamma \beta, \label {onsager} \ee where $\gamma$ is the
symmetric matrix describing the transition rates at equilibrium,
defined earlier. This puts the results of the linearization
procedure of [\onlinecite{amir_glass}] in a much more natural
context, and explains why the matrix $A$ describing the relaxation
is not hermitian in the general case: the multiplication of the
symmetric matrix $\gamma$ and the symmetric matrix $\beta$ is not
expected to be (and is generally not) a symmetric matrix.

\subsection {Langevin approach}
\label{langevin}

Due to the finite temperature, we expect the system to fluctuate
around the equilibrium distribution. This can be modeled using a
Langevin equation.  In a similar fashion to the usual treatment of
Brownian motion, where the equation governing the motion is $
\frac{d\vec{v}}{dt}=-\frac{\vec{v}}{\tau}+\vec{f}(t) , \label
{brownian}$ we add a Langevin noise term to  Eq. (\ref{dynamics}):

\be \frac{d \vec{\delta n}}{dt}= A \cdot \vec{\delta n} +
\vec{f}(t),\label {modified}\ee where $f_i$ is a white-noise term.
We shall now show, however, that the different components of the
noise vector must be correlated, i.e., $\langle f_i f_j\rangle
\equiv F$ is a non-diagonal matrix. As in the case of Brownian
motion, thermodynamic considerations determine the magnitude of the
noise, as well as the correlations between the different vector
components. It is known that $F=A\beta^{-1}=\gamma$ \cite{LL}. There
is a clear physical intuition for this: correlations between sites
$i$ and $j$ are contributed from the direct microscopic current
between the two sites, $\gamma_{ij}$. Since $\gamma$ is not
diagonal, there exist correlations between the different components
of the noise vector, as stated previously. See the Appendix for an
explanation of the physical meaning of the matrix elements of
$\gamma$ as kinetic coefficients.

%
%The matrix $\gamma$ has the physical meaning of kinetic
%coefficients. Using Eq. \ref{mu} we see that:
%
%\be \frac{d\vec{n}}{dt}=A \beta^{-1} \vec{\mu}=\gamma \vec{\mu},\ee
%therefore the elements of $\gamma$ describe the response of the
%'fluxes' to the 'forces', justifying the term kinetic coefficients.

\subsection {Form of the spectrum}
\label{formofspectrum} It is useful to diagonalize
Eq.~(\ref{eq:phi}) by multiplying both sides on the left by
$\beta^{1/2}_{mi}$. We then have, using (\ref{onsager}) and a matrix
notation:
\begin{equation}
\label{eq:phi1} \frac {d(\beta^{1/2}\phi(t))}{dt}=\beta^{1/2} \gamma
\beta^{1/2} (\beta^{1/2} \phi(t)).
\end{equation}

Defining $\tilde \phi = \beta^{1/2}\phi$ and $\tilde \gamma =
\beta^{1/2} \gamma \beta^{1/2}$ we reduce the equation to a
symmetric form (notice that $\tilde \gamma = \tilde \gamma^T$):
\begin{equation}
\label{eq:phitilde} \frac {d\tilde \phi(t)}{dt}=\tilde \gamma \tilde
\phi(t).
\end{equation}
We can now diagonalize the symmetric operator $\tilde \gamma$ in a
standard way:
\begin{equation}
\tilde u^\dagger  \tilde \gamma \tilde u = \tilde \Lambda
\end{equation}
with $\tilde \Lambda$ a diagonal matrix, and the elements of $\tilde
u_{ij} = \psi_i^j$ are the orthonormal eigenbasis ($\tilde u^\dagger
\tilde u= \tilde u \tilde u^\dagger=I$) of $\tilde \gamma$. The
eigenvalues of $\tilde\gamma$ turn out to be identical to those of
$A$. Notice that this proves that the eigenvalues of $A$ are real.
To see this, it is useful to utilize the definition $u^{-1} \equiv
\tilde u^\dagger \beta^{1/2}, u =\beta^{-1/2} \tilde u \Rightarrow
\tilde u = \beta^{1/2} u$. Thus:
\begin{equation}
\label{eq:lambdatildelambda} \tilde \Lambda = \tilde u^\dagger
\beta^{1/2} \gamma \beta^{1/2} \tilde u = u^{-1} \gamma \beta^{1/2}
\tilde u = u^{-1} \gamma \beta u =  u^{-1} A u
\end{equation}

We can immediately write the solution for $\tilde \phi$:
$$
\tilde \phi = e^{-\tilde \gamma t} \tilde \phi(t=0) = \tilde u
e^{-\tilde \Lambda t} \tilde u^\dagger \tilde \phi(t=0) \Rightarrow
\phi =\beta^{-1/2} \tilde u e^{-\tilde \Lambda t} \tilde u^\dagger
\beta^{-1/2} \label {correlations}
$$
%
%\be \phi =\beta^{-1/2} \tilde u e^{-\tilde \Lambda t} \tilde
%u^\dagger \beta^{-1/2} \label {correlations} \ee

 Explicitly: \be \phi_{ii}=\sum
\beta^{-1/2}_{ij}\psi^{*\alpha}_j e^{-\lambda_\alpha t}
\psi^\alpha_k \beta^{-1/2}_{ki}. \label{noise_numeric_eq}\ee

This equation is an exact formula for the correlation function at
site $i$, and will be later used to numerically find the spectrum.
At low temperatures, as mentioned in section \ref{Equilibrium
correlations}, the occupations tend to 0 and 1 exponentially since
they follow Fermi-Dirac statistics. For all sites whose distance
from the Fermi-energy is larger than the temperature, we can
therefore neglect the off-diagonal elements of the matrix. This is
nevertheless a non-trivial approximation, since the sites close to
the Fermi-energy are the ones contributing the most to the noise.
For this reason we test the validity of this approximation
numerically, and show that it doesn't change the form of the
spectrum at the end of this section. Making the approximation we
obtain:
$$
\phi_{ii}(t) = \sum_{\alpha} \frac{1}{\beta_{ii}}
\left|\psi^\alpha_i\right|^2 e^{-\lambda_\alpha t}
$$

Or in the frequency domain:

\begin{equation}
\label{eq:sumform}
 \phi_{ii}(\omega) = \sum_{\alpha}
\frac{1}{\beta_{ii}} \left|\psi^\alpha_i\right|^2 \frac{2
\lambda_\alpha}{\omega^2+\lambda_\alpha^2}
\end{equation}

This is an explicit formula for the noise spectrum. We can make a
further approximation, by noticing that for most sites
$1/\beta_{ii}= n_i (1-n_i)$ is very close to zero, since $n_i$ is
the Fermi function of the on-site energy, which is typically much
greater than the temperature. Therefore, only a small fraction of
sites for which the on-site energy (renormalized by the interaction)
is of the order of $kT$ away from the Fermi energy, will contribute
to the sum. This is physically clear: it is exactly those sites
which are close to the Fermi energy whose charge can fluctuate and
contribute to the noise. Other sites are nearly permanently empty or
full. For sites with energies $E_i$ a distance \emph{much} smaller
than $kT$ from the Fermi energy, $n_i (1-n_i) = 1/4$. We therefore
make an approximation and sum only over a partial number of the
sites, for which we replace $1/\beta_{i,i}$ by $1/4$. This leads us
to the equation:

\begin{equation}
 \phi_{ii}(\omega) \sim \sum_{\alpha}
\frac{1}{4} \left|\psi^\alpha_i\right|^2 \frac{2
\lambda_\alpha}{\omega^2+\lambda_\alpha^2}
\end{equation}

Taking another average over the different sites, we can replace
$\langle \langle(\psi^\alpha_i)^2\rangle \rangle$ by a constant
$\frac{1}{N}$, leading to:

\be \langle \langle \delta n^2\rangle \rangle_\omega  \sim
\frac{1}{N}\sum_{\alpha,i}
  \frac{\frac{2}{\lambda_\alpha}}{1+(\frac{\omega}{\lambda_\alpha})^2} , \label {t_dep}\ee
  where $\langle \langle;\rangle \rangle$ denotes averaging over sites as well as time.

At this stage the equation takes the form leading to $1/f$ noise in
various other theories: a discrete sum of equally contributing
Lorentzians. The crucial thing is that the non-trivial weights
arising from the matrix $\beta$ did not affect the structure of the
noise statistics. Finally, turning the sum into an integral and
using the fact that $P(\lambda) \sim \frac{1}{\lambda}$ in a large
window \cite{amir_glass}, we obtain:

\be \langle \langle \delta n^2\rangle \rangle_\omega  \sim
\frac{1}{N}\int_{\lambda_{min}}^{\lambda_{max}} d\lambda
  \frac{\frac{1}{\lambda ^2}}{1+(\frac{\omega}{\lambda})^2} =\frac{1}{N
\omega}\int_{\frac{\lambda_{min}}{\omega}}^{\frac{\lambda_{max}}{\omega}}
dm
  \frac{1}{1+m^2} . \label {w_dep}\ee

%\be \langle \langle\delta n^2\rangle \rangle_w  = \frac{1}{N
%\omega}\int_{\frac{\lambda_{min}}{\omega}}^{\frac{\lambda_{max}}{\omega}}
%dm
%  \frac{1}{1+m^2} .\ee
This shows that for $\lambda_{min} \ll \omega \ll \lambda_{max}$, a
\emph{1/f} spectrum indeed follows for the noise in the average
occupation number, which is one of our main results. In the future
we intend to study the implications of the fluctuations in the
occupation numbers on the conductance fluctuations.

\begin{figure}[t!]
\includegraphics[width=0.7\textwidth]{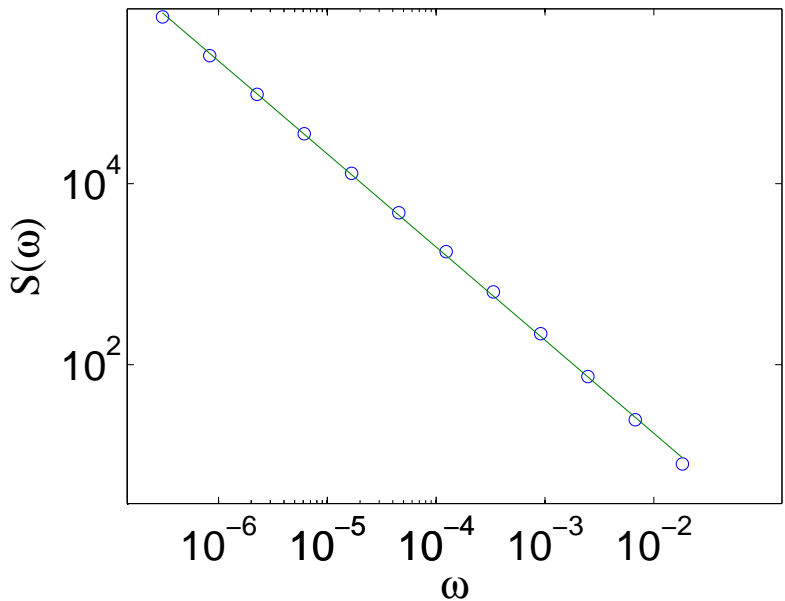}
\caption{A numerical test of the noise spectrum. In the first step,
the equilibrium occupation numbers, energies and transition rates
were found for an arbitrary metastable state, using the procedure
descrived in [\onlinecite{amir_glass}]. Next, the matrix $\beta$ was
constructed using Eq. (\ref{beta}), and the spectrum of $A$ defined
in Eq. (\ref{onsager}) was found. Finally, the spectrum was computed
using the Fourier transform of Eq. (\ref{noise_numeric_eq}), and
averaged over the different sites. The spectrum was averaged over
100 realizations. The best fit over more than 4 decades gave a slope
of $-1.03 \pm 0.03$. The system is one-dimensional, with $N=100$
sites, and $20T= e^2/r_{nn}=W$, where $T$ is the temperature,
$r_{nn}$ the average nearest-neighbor distance, and $W$ the
magnitude of the disorder. $\xi/r_{nn}=0.1$, where $\xi$ is the
localization length. \label{numerics} }
\end{figure}

We have performed a numerical test of the calculation, taking the
complete form of the matrix $\beta$, including the off-diagonal
elements, and using Eq. (\ref{noise_numeric_eq}). Figure
\ref{numerics} shows the results for the spectrum, as well as
explanations of the numerical procedure. The best linear fit over
more than 4 decades gave a slope of $-1.03 \pm 0.03$, close to the
expected result.

\section {A proposal for an experiment}

A \emph{1/f} noise spectrum has been observed experimentally in
electron glasses \cite{massey, mccammon, kar, zvi_exp1}. The
measurements performed so far, however, were always for the
conductance. This is not the quantity we calculate in this work,
although [\onlinecite{amir_aging}] shows a strong connection between
the excess conductance and the deviations of the occupations. We
hereby propose another possible experiment, in which the
fluctuations in the occupation numbers are measured directly. In
this experiment a small capacitor is placed in close proximity to
the sample. The fluctuations in the occupation numbers of the
neighboring sites would induce fluctuations in the voltage of the
capacitor, which can be measured. Averaged over the different
location in the sample, these should obey the \emph{1/f} noise
discussed in section \ref{sect4}. A similar experimental technique
has been implemented by [\onlinecite{ashoori, eisenstein, yacoby}].
It would be very interesting to see whether a change in the nature
of the noise is seen across the metal-insulator transition in such a
measurement.

%\clearpage

%generated by noise_LL_spectrum_1d.m

%\begin{acknowledgement}
 This work was supported by a BMBF
DIP grant as well as by ISF and BSF grants and the Center of
Excellence Program. A.A. acknowledges funding by the Israel Ministry
of Science and Technology via the Eshkol scholarship program.

%\end{acknowledgement}

\appendix \section {Onsager symmetry of kinetic coefficients}

\normalsize It is the purpose of this Appendix to give a concise
derivation of the Onsager regression hypothesis and the Onsager
symmetry of the kinetic coefficients, along the lines of
[\onlinecite{LL}].

 Let us consider fluctuations of variables $n_i$ around some stable state in
thermal equilibrium. For convenience we take the equilibrium values
to be 0. Close to the stable point we can expand the free energy to
second order (the first order vanishes since we are at an extremum),
and arrive at Eq. (\ref{Fexpand}):

\be F= \sum_{i,j} \frac{kT}{2} \beta_{ij} n_i n_j. \ee

The Boltzmann distribution immediately gives us, after a gaussian
integration, that:

\be \langle n_i(0) n_j(0) \rangle = \beta^{-1}_{ij}, \ee

where $\langle \rangle$ is the statistical ensemble average. We see
that the inverse of $\beta$ gives the equal-time correlation matrix.

We now define the conjugate variable to $n$ (the thermodynamic
forces), as in section \ref{sect4}:

\be \mu_i=\frac{\partial F}{\partial n_i} = \beta_{ik}n_k. \ee

A similar calculation to the one done for the  correlations of $n$
gives that $\langle \mu_i \mu_j \rangle = \beta_{ij}$, and $\langle
n_i \mu_j \rangle = \delta_{ij}$.

Let us assume the linearized equation of motion at the stable state
is $\dot{\vec{n}}=A\vec{n}.$

Defining $\phi_{ij} \equiv \langle n_i(t) n_j(0) \rangle$, we obtain
that:

\be \dot{\phi}_{ij} = \langle \dot{n_i}n_j(0)\rangle=A_{ik}\langle
n_k(t)n_j(0)\rangle. \ee Thus we have the matrix equation: \be
\dot{\phi}=A \phi. \ee This is the \emph{ Onsager regression
principle}, stating that the correlation function obeys the same
relaxation equation as that of the microscopic variables.

Let us look at $C_{ij}(t) \equiv
\frac{d\phi_{ij}}{dt}=\frac{d\langle {n_i(t)}n_j(0) \rangle}{dt}$.
Assuming time-reversal symmetry , $C$ is a symmetric matrix: Since
we are at equilibrium we have time translation invariance, thus
$\frac{d\langle n_i(t)n_j(0) \rangle}{dt}=\frac{d\langle
n_i(0)n_j(-t) \rangle}{dt}$, and using the time-reversal symmetry
this is equal to $\frac{d \langle n_i(0)n_j(t)\rangle}{dt}=C_{ji}$.
Taking $t=0$ we obtain:

\be C_{ij}(0)=\langle A_{i,k}n_k(0)n_j(0)\rangle =
A_{i,k}\beta^{-1}_{kj}. \ee

Thus, $A \beta^{-1} \equiv \gamma$ is a symmetric matrix.

Notice that $\dot{\vec{n}}=A \beta^{-1} \beta \vec{n} = \gamma
\vec{\mu}$, which is why the elements of $\gamma$ are called kinetic
coefficients, relating the fluxes (derivatives of $n$) to the forces
$\mu$.

\end{document}